\documentclass[%
 aps,
 pra,%
longbibliography,
 amsmath,amssymb,
 reprint,%
superscriptaddress,
]{revtex4-1}

\usepackage{graphicx}
\usepackage{bm}
\usepackage{times}
\usepackage{amsmath}
\usepackage{bbold}
\usepackage[english]{babel}
\usepackage{color}
\usepackage[colorlinks]{hyperref}
\hypersetup{
colorlinks=true,
linkcolor=blue,          
citecolor=blue,          
filecolor=magenta,       
urlcolor=cyan
}

\begin{document}

\title{Sub-pico-liter magneto-optical cavities}

\author{J. A. Haigh}
	\email{jh877@cam.ac.uk}
\affiliation{Hitachi Cambridge Laboratory, Cambridge, CB3 0HE, United Kingdom}

\author{R. A. Chakalov}
\affiliation{Cavendish Laboratory, University of Cambridge, Cambridge, CB3 0HE, United Kingdom}

\author{A. J. Ramsay}
\affiliation{Hitachi Cambridge Laboratory, Cambridge, CB3 0HE, United Kingdom}

\date{\today}

\begin{abstract}
Microwave-to-optical conversion via ferromagnetic magnons has so-far been limited by the optical coupling rates achieved in mm-scale whispering gallery mode resonators. Towards overcoming this limitation, we propose and demonstrate an open magneto-optical cavity containing a thin-film of yttrium iron garnet (YIG). We achieve a 0.1\,pL (100\,$\mu$m$^{3}$) optical mode volume, $\sim$50 times smaller than previous devices. From this, we estimate the magnon single-photon coupling rate is $G\approx50$\,Hz. This open cavity design offers the prospect of wavelength scale mode volumes, small polarization splittings, and good magneto-optical mode overlap. With achievable further improvements and optimization, efficient microwave-optical conversion and magnon cooling devices become a realistic possibility.
\end{abstract}

\maketitle

\section{Introduction}

Magnetic-field tunable ferromagnetic modes can be easily strongly coupled to microwave resonators  \cite{tabuchi_hybridizing_2014,huebl_high_2013-1,zhang_strongly_2014-1}. Further coupling to optical photons offers the prospect of useful transduction of microwave quantum signals to telecoms optical wavelengths \cite{hisatomi_bidirectional_2016}. For this reason, the interaction of magnons and optical photons has been explored recently in the whispering gallery modes (WGM) of yttrium iron garnet (YIG) spheres \cite{osada_cavity_2016,zhang_optomagnonic_2016,haigh_triple-resonant_2016}. However, despite the high $Q$-factor of the magnetic and optical modes \cite{haigh_magneto-optical_2015}, the optomagnonic coupling rates achieved in mm-scale YIG spheres have been limited to $\sim$1\,Hz. If the coupling rate can be increased significantly, in turn raising the conversion efficiency, this would open a wide range of technological opportunities \cite{lambert_coherent_2020}, as well as the ability to coherently modify the magnetization dynamics, for example cooling or dynamical driving the magnon mode \cite{sharma_optical_2018,simic_coherent_2020}.

The low coupling rate for optical whispering gallery modes is due to the poor mode overlap and the large volume of magnetic material involved. To overcome the poor overlap, it may be possible to exploit \emph{magnon} whispering gallery modes in YIG spheres, with almost ideal overlap with the optical WGM \cite{sharma_optimal_2019}. A simpler strategy is to explore more compact structures, as very recently shown in rib waveguide devices \cite{zhu_waveguide_2020}. In that case, the mm-long structure confines both the magnons and photons, yielding excellent overlap inside the structure, enabling a coupling rate of 17\,Hz.

The estimated maximum coupling rate for a YIG optical resonator is $\approx 0.1$\,MHz \cite{viola_kusminskiy_coupled_2016-1}, based on a mode volume of the order of the resonant wavelength cubed $\lambda^3$. There are several candidate wavelength-scale optical resonators \cite{vahala_optical_2003}, which could get  close to this maximum coupling rate. The choice of resonator, however, must take into account the significant challenges of micro-patterning YIG \cite{heyroth_monocrystalline_2019}. We note that, while sub-wavelength mode confinement is possible with plasmonic devices \cite{yu_plasmon-enhanced_2019,schlauderer_temporal_2019,maccaferri_nanoscale_2020}, this typically comes with high optical losses in the metal components.

A simple optical resonator design, with wavelength scale mode volumes combined with large $Q$-factors, is an \emph{open microcavity} \cite{trupke_microfabricated_2005}. These are typically hemispherical resonators where a reflection-coated microlens is positioned in close proximity to a mirror surface. Devices can be fiber-based \cite{steinmetz_stable_2006,muller_coupling_2009}, or fabricated on planar surfaces \cite{cui_hemispherical_2006}, and have previously been used to obtain large coupling rates to single atoms \cite{trupke_atom_2007}, N-V centers \cite{albrecht_coupling_2013}, single organic dye molecules \cite{wang_coherent_2017}, and excitons in 2D materials \cite{dufferwiel_excitonpolaritons_2015}. The advantage of this structure is that any transferable material can be easily embedded \cite{barbour_tunable_2011}, and the modes are tunable by the position of the lens. Optical mode volumes as small as 1\,fL (1\,$\mu$m$^{3}$) have been achieved, with $Q$-factors in excess of 10,000 \cite{dolan_femtoliter_2010}.

In this article, we demonstrate a viable route to low mode volume, high coupling rate magneto-optical cavities with mode volumes limited by the optical wavelength. We embed single crystal YIG layers in an open microcavity, and show a two orders of magnitude increase in the coupling rate over whispering gallery mode devices. With further reduction in mode volume, it is expected that the strong-coupling limited can be reached. This work, therefore, shows a path towards efficient microwave-optical conversion, and optical magnon cooling.

\section{Coupling rate}

We first briefly review the enhanced scattering process in cavity optomagnonics. Magnetic Brillouin light scattering is an inelastic process where a photon is scattered from an input mode $\hat{a}_i$ into an output mode $\hat{a}_o$, with absorption or emission of a magnon in mode $\hat{m}$. Brillouin light scattering is most efficient between orthogonally polarized optical modes, as this compensates the angular momentum lost or gained to the magnon mode, conserving angular momentum. To enhance the BLS significantly, we require two optical resonances, enhancing both the input and output optical fields. These should be orthogonally polarized, and with frequency separation matching the magnon frequency. This is the \emph{triple resonance condition}, which has been observed previously in whispering gallery mode resonators \cite{haigh_triple-resonant_2016}, and in a recent waveguide device \cite{zhu_waveguide_2020}.

The interaction Hamiltonian that governs the scattering is of the form $H_\text{int} = G ( \hat{a}^{\dagger}_i \hat{a}^{}_o \hat{m} + \hat{a}^{}_i \hat{a}^{\dagger}_o \hat{m}) $, with interaction strength quantified by the coupling rate,
%
%
\begin{equation}
G = -i  \theta_f \frac{c}{n} \sqrt{\frac{4 g \mu_B}{M_s}} \sqrt{\frac{\eta_\text{mag}\eta_\text{opt}}{V_\text{opt}}}.
\label{couplingrate}
\end{equation}
The numerical constant $\mu_B$ is the Bohr magneton and $c$ the speed of light. The factors affecting this rate can be separated in two parts. Firstly, the materials parameters of the embedded magnetic material: the Faraday coefficient $\theta_f$, refractive index $n$, gyromagnetic ratio $g$ and saturation magnetization $M_s$. These parameters can be optimized by materials development, finding new materials and improving the quality of those available. Secondly, the geometry of the optical cavity affects the coupling rate through the volumes of the optical modes $V_i \approx V_o = V_\text{opt} = \int{|\mathbf{u}_{i,o}(\mathbf{r})|^2}$, where $\mathbf{u}_{i,o}(\mathbf{r})$ is the mode function with normalization $\max{(|\mathbf{u}_{i,o}(\mathbf{r})|^2)}=1$. The overlap is contained in the fill-factors $\eta_\text{mag} = V_\text{int}/V_\text{mag}$ and $\eta_\text{opt} = V_\text{int}/V_\text{opt}$, which are the proportion of the magnetic $V_\text{mag}$ and optical $V_\text{opt}$ modes volumes that contribute to the coupling through the triple-mode overlap,
\begin{equation}
V_\text{int} = \int{d\mathbf{r}\,\mathbf{u}_{m}(\mathbf{r}) \cdot[\mathbf{u}_i^{*}(\mathbf{r})\times\mathbf{u}_o(\mathbf{r})]}.
\label{Vint}
\end{equation}
Here, the mode function $\mathbf{u}_{m}$ is also normalized such that $\max{(|\mathbf{u}_m(\mathbf{r})|^2)}= 1$. This expression includes the effect of the mode polarization. To maximize the geometric factors we would like a low optical mode volume resonator, with excellent overlap with the magnon mode, and orthogonal polarization of all three modes.

In this paper, we focus exclusively on the minimization of the optical mode volume. The design is such that lateral patterning of the continuous YIG layer to confine the magnon mode can incorporated at a later stage.

\section{Design}

\begin{figure}%
\includegraphics[width=\columnwidth]{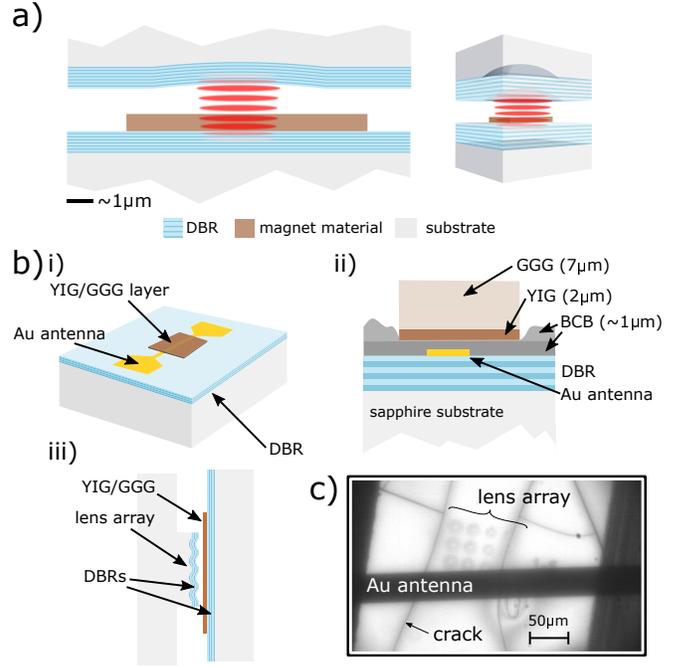}
\caption{(a) Schematic of the magneto-optical cavity design. The open microcavity consists of two parts: a concave lens milled into a substrate and coated with a DBR, and a planar mirror with magnetic layer. Left: Cross section of design to show relative dimensions of the lens and beam waist. Right:  3D representation of the device structure. (b)Fabrication of magneto-optical microcavity. (i) Flat side of open microcavity. A gold microwave antenna is patterned on the DBR surface before the YIG/GGG layer is bonded. (ii) Cross section of flat mirror, showing two-layer BCB bonding polymer. (iii) Cross-section section of open microcavity, showing microlens array. (c) Optical image through cavity structure, showing lens array and microwave antenna.}%
\label{design}%
\end{figure}

A schematic of our proposed device is shown in Fig.\,\ref{design}(a). The mirrors forming the open microcavities are purchased from Oxford HiQ \footnote{Oxford HiQ \url{https://www.oxfordhighq.com/}.}. They consist of one planar surface and one microlens array. Both surfaces have a high quality reflective coating consisting of 22 layers of SiO$_2$ and Ta$_2$O$_5$, deposited by sputtering. This distributed Bragg reflector (DBR) has a reflectivity of ~99.8$\%$ at its design wavelength of 1300\,nm. The microlens array contains 16 lenses with four different radii of curvature, from 100\,$\mu$m down to 20\,$\mu$m, fabricated by focused-ion-beam milling \cite{trichet_topographic_2015}. The lens array is situated on a raised pedestal to allow alignment of the two mirrors.

To embed a high quality, single crystal YIG layer in the microcavity, we avoid deposition techniques such as pulsed laser deposition and sputtering, because the non-lattice matched DBR substrate would lead to poly-crystalline YIG growth \cite{goto_vacuum_2013}. Furthermore, post growth annealing to improve the crystallinity of those layers requires temperatures above 700\,$^\circ$C, which has been found to be detrimental to the DBR \cite{yoshimoto_magnetophotonic_2016}. Instead, we use a lift-off technique \cite{levy_epitaxial_1997} to remove a layer of single-crystal YIG from a lattice matched GGG growth substrate. We later bond this layer to the mirror surface with a spin-on polymer.

The advantage of the open microcavity design is that the polarization splitting is minimized due to the cylindrical symmetry. In principle, a minor asymmetry can tune the splitting to match the magnon frequency, given the precise control of lens profile that has been demonstrated \cite{trichet_topographic_2015}. For an optical resonator with an asymmetrical cross-section, the splitting is typically a fixed fraction of the free spectral range. Therefore, as the mode volume shrinks, the frequency separation can become too large. This can be seen in the whispering gallery mode resonators, where a 1\,mm diameter YIG sphere was chosen to match the splitting to the magnon frequency \cite{haigh_triple-resonant_2016}. It is not possible to decrease the size of the sphere, because the splitting would become too large. This effect can also be seen in the rib waveguide geometry \cite{zhu_waveguide_2020}, where the length of the cavity must be long ($\approx4$\,mm) to keep the polarization splitting small.


\section{Fabrication}

We start with a YIG film of thickness 2\,$\mu$m grown by liquid phase epitaxy on a gadolinium gallium garnet (GGG) substrate \footnote{Innovent Jena \url{https://www.innovent-jena.de/}.}. The lift-off is achieved by inducing spontaneous delamination as follows. The sample is first subjected to a high dose ion implantation $5\times10^{16}$\,cm$^{-2}$ with He ions at 3.5\,MeV \footnote{Ion implantation was carried out at University of Surrey Ion Beam Centre.}. The penetration depth is approximately 9\,$\mu$m, with straggle $\approx$1\,$\mu$m, creating a narrow layer in which the lattice is substantially damaged \cite{levy_crystal_1998}. Annealing at 470\,$^{\circ}$C for 1\,min leads to  delamination of a bilayer consisting of the 2\,$\mu$m of YIG and around 7\,$\mu$m of GGG . This delamination occurs due to the slight lattice mismatch and different coefficient of expansion of YIG and GGG \cite{levy_crystal_2000}. The lattice mismatch leads to a  membrane with $\sim$10-mm radius of curvature  at room temperature.

The YIG wafer is diced into 1\,mm square chips post implantation, but prior to delamination.
%
Post delamination, the thin membrane is manipulated using a small piece of 25\,$\mu$m thick Kapton film, where it is held in place by static.

To bond the YIG/GGG membrane to the mirror surface, we use BCB cyclotene \cite{noauthor_cyclotene_2012}, a polymer used as a dielectric in microelectronics, adhesive wafer bonding \cite{niklaus_low-temperature_2001} and planarization applications \cite{stokich_planarization_1993}. It has excellent optical properties \cite{zhao_polymer_2000}, allowing its use, for example, in bonding active III-V devices to silicon photonic wafers \cite{roelkens_adhesive_2006}.

Prior to bonding, a strip-line antenna is patterned on to the surface of the mirror using photo-lithography and lift-off as shown in Fig.\,\ref{design}(b). A titanium adhesion layer of 7\,nm is deposited under 100\,nm of gold, with a final 7\,nm of titanium above. This final layer is required to avoid the poor adhesion of BCB cyclotene to gold \cite{noauthor_processing_2007}.

We prepare the mirror surface with solvent cleaning in an ultrasonic bath. The device is then soaked in DI water, before a 2\,min plasma cleaning process in a reactive ion etcher. This is followed by a further 2\,min soak in DI water. The surface is primed with an adhesion promoter AP3000 \cite{noauthor_processing_2007}. The BCB cyclotene is deposited and spun for 30s at 6000 RPM, followed by 1\,min on a hot plate at 150\,$^{\circ}$C to remove the solvent. We use a double layer of BCB cyclotene \cite{roelkens_adhesive_2006}. The first layer is partially cured with a 2\,min anneal at 250\,$^{\circ}$C on a strip annealer. This layer remains `tacky' and bonds well to a second layer of BCB cyclotene, but is viscous enough to prevent pinch-through under the membrane during curing, where there can be significant re-flow of the polymer \cite{roelkens_adhesive_2006}. The second layer of BCB is spun under the same conditions.

Separately, the YIG/GGG layer is prepared with a 2\,min plasma ash to activate the surface, before a 12\,hr evaporation of AP3000 is performed in a desiccator. The membrane is removed from the desiccator immediately prior to bonding.

The bonding is performed in a simple spring-loaded clamp. The Kapton tape bearing the YIG/GGG membrane is placed on the mirror, with the YIG layer in contact with the BCB cyclotene. The clamp is closed to the point where the layer is held in place with minimal pressure and then heated on a hot plate to 150\,$^{\circ}$C. The pressure is then increased to the required load. The assembly is then transferred to an oven at 150\,$^{\circ}$C under nitrogen flow to prevent oxidation of the BCB cyclotene at elevated temperatures. The oven temperature is ramped to 250\,$^{\circ}$C at 1\,$^{\circ}$/min, for a 1\,hr soak. After allowing the oven to cool to room temperature, the clamp is removed and the Kapton film peeled from the mirror surface. This leaves the YIG/GGG layer secured to the device.

During the bonding process, there is some re-flow of the BCB cyclotene to the top surface of the YIG/GGG membrane. To remove this, a 3\,min Ar/CF$_4$ reactive ion etch descum is performed.

\section{Experimental setup}

For measurement, the planar mirror is glued over an aperture on a PCB patterned with input and output coplanar waveguides, which are connected to semi-rigid coaxial cables. The on-chip strip-line antenna is then wire-bonded to the PCB waveguides for microwave measurement and excitation of the magnon modes in the YIG. The PCB is mounted on a circular stub which sits in an $xyz$-translation lens mount. The lens array is similarly mounted on a circular stub in a tilt-yaw lens holder, for full control of the cavity geometry.

\begin{figure}%
\includegraphics[width=\columnwidth]{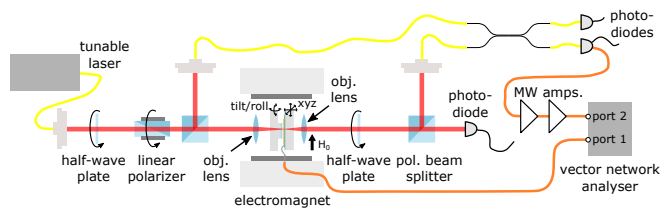}%
\caption{Experimental setup. The output polarization of a 1270-1370\,nm tunable laser is controlled via a linear polarizer and a half-wave plate, before being separated into a local oscillator and cavity drive. After passing through the cavity, the optical signal orthogonal to the input polarization is recombined with the local oscillator and measured on a high frequency photodiode. The transmission through the cavity is measured via the light with the same polarization as the input on a dc photodiode. A vector network analyzer drives the magnetic modes and measures the microwave signal from the fast photodiode.}%
\label{setup}%
\end{figure}

The device is mounted in an electromagnet, with magnetic field applied orthogonal to the cavity length. Light is focused into and out-of the cavity using two aspheric lenses mounted on $xyz$ stages. The cavity is selected by scanning the laser to the correct position. The input laser is an external cavity diode laser with linewidth $\approx$1\,MHz. The input polarization is set with a rotatable Glan-Thompson prism. On the output, a rotatable half-wave plate is used to select the measurement basis on a polarizing beam splitter. From the beam splitter, the transmitted signal with the same polarization as the input light field is measured with a dc photodiode. The polarization scattered light is focused into a single mode fiber, and combined with a local oscillator directly from the laser in a 50:50 fiber coupler. One output of this coupler is measured on a fast photodiode (12\,GHz bandwidth) connected via a microwave amplifier to a vector network analyzer (VNA). The VNA is also used to drive the magnetization dynamics via the microwave antenna.

\section{Characterization}\label{sec:char}

\begin{figure}%
\includegraphics[width=\columnwidth]{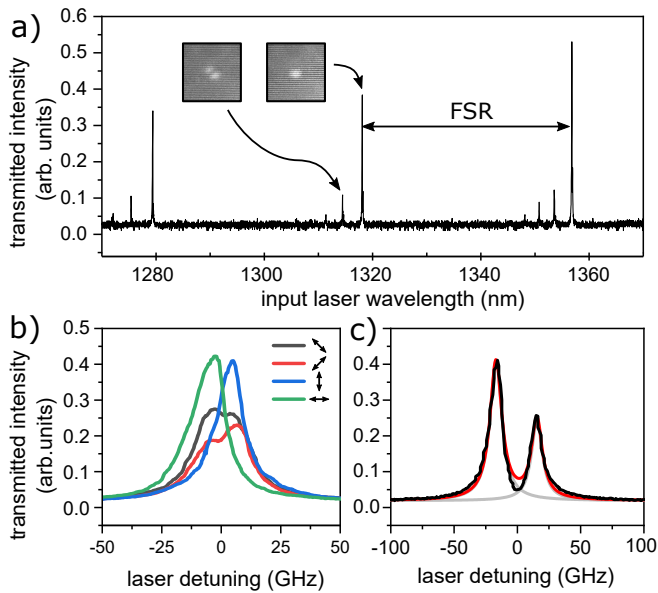}%
\caption{Transmission spectroscopy of optical modes. (a) Wide wavelength scan, showing free spectral range. Insets show mode profile imaged in transmission. (b) Measurement of polarization of modes. The linear polarization can be set so that only one mode is excited. This device had a smaller polarization splitting $\approx$16\,GHz. (c) Measurement of polarization splitting and optical linewidth of device used in BLS measurements. The linear polarization is set so that both modes are probed. This device is also measured in (a).}%
\label{modes}%
\end{figure}

We first characterize the optical modes of the microcavities. The transmitted intensity is measured as a function of input laser wavelength, and angle of input linear polarization, as shown in Fig.\,\ref{modes}. A measurement over a wide wavelength range (Fig.\,\ref{modes}(a)) is used to determine the free spectral range $\Delta\omega_\textsc{fsr}/2\pi\approx6.7$\,THz. A number of spatial modes resulting from the lateral confinement of the microlens are visible. These can be identified by imaging in transmission, see insets of Fig.\,\ref{modes}(a). The coupling to these higher order modes is minimized by optimizing the transmitted intensity on resonance through the lowest order mode.

By measuring the transmitted intensity as a function of the angle of linear polarization, we can find the axes of the orthogonal, linearly polarized modes, and the splitting between the two. An example of this measurement is shown in Fig.\,\ref{modes}(b), where the polarization splitting is 16\,GHz. This splitting varies with different lens arrays, and is related to slight asymmetries in the nominally-cylindrical fabricated lens. In the device used for BLS measurements shown in this paper, the splitting is 32\,GHz, as shown in Fig.\,\ref{modes}(c). Because the applied magnetic field from the electromagnet is limited to $<$1\,T, we are unable to reach the triple resonance condition. We note that the  frequency splitting due to the magnetic linear birefringence in YIG \cite{pisarev_magnetic_1971} is estimated as $\sim$900\,MHz. This is not large enough to explain the observed splittings. We note that by fabricating arrays of lenses with varying ellipticity, it would be possible to obtain microcavities with a specific splitting. This would enable the triple resonance condition to be achieved.

We extract the total dissipation of the optical mode from the linewidth of peak (Fig.\,\ref{modes}(c)) $\kappa_{}/2\pi=11$\,GHz. This corresponds to a Q-factor of 20,000 and Finesse of 600.

The expected external loss rate can be estimated from the DBR reflectivity $R=0.9986$ as $\kappa_\text{ext} = - 2 \Delta\omega_\textsc{fsr} \log{R}$ \cite{ismail_fabry-perot_2016}, giving $\kappa_\text{ex}/2\pi \approx$3\,GHz. Using these values, and $\kappa_{}=\kappa_\text{ext} + \kappa_\text{int}$, we can estimate the internal dissipation rate $\kappa_\text{int}/2\pi=$8\,GHz. This is consistent with the transmitted intensity on resonance $\kappa_\text{ext}^2/\kappa^2\approx0.07$. If this internal dissipation were solely due to absorption in the YIG layer, we would expect $\kappa_\text{int}=\kappa_\text{abs} = (\alpha c / n_\textsc{yig}) (t_\textsc{yig}/L)\approx1$\,GHz. The discrepancy suggests that other dissipation mechanisms play a role. A likely source is the surface roughness on the GGG top surface, where the crack propagates during lift-off. This could be alleviated by post-bonding polishing.

The choice of mirror reflectivity  was conservative to ensure good coupling to the cavity. If the scattering losses can be eliminated, then the mirror reflectivity could be increased, while keeping the system over-coupled.  In this case, the minimum possible dissipation rate would be $\kappa_\text{abs}\sim1$\,GHz - as achieved in WGM cavities \cite{zhang_optomagnonic_2016,haigh_magneto-optical_2015}.

\section{Brillouin light scattering}\label{sec:BLS}

Next, we use homodyne detection to measure the magnon-scattered light, emitted from the microcavity with opposite linear polarization to the input. The input laser wavelength is fixed and set to the lower wavelength optical mode, with frequency $\omega_i$. The VNA is used to drive the magnon modes via the microwave antenna, as well as detect the signal at the same frequency from the fast photodiode, where the scattered light is combined with a local oscillator taken from the input laser.

\begin{figure}%
\includegraphics[width=\columnwidth]{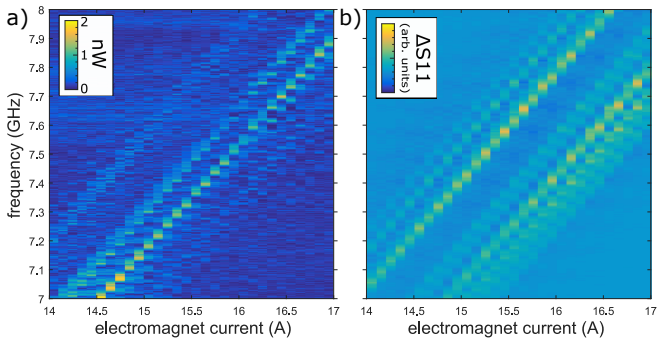}%
\caption{(a) Brillouin light scattering signal. The mixed power with the local oscillator, incident on the fast photodiode. A magnetic field independent background has been subtracted. (b) Microwave measurement of magnetic modes, via $|S11|$ using the vector network analyzer.}%
\label{BLS}%
\end{figure}

A measurement using this method is shown in Fig.\,\ref{BLS}(a), as a function of microwave drive frequency and applied magnetic field. When the microwave drive is resonant with a magnon mode with frequency $\omega_m$ within the microcavity, the magnons created scatter with the input optical photons to create optical photons at a frequency $\omega_i\pm\omega_m$. When combined with the local oscillator $\omega_{LO} = \omega_i$, and mixed on the photodiode, this results in a microwave signal at $\pm\omega_m$, resulting in the bright lines in Fig.\,\ref{BLS}(a). The power plotted is the optical power at the photodiode, using the responsivity of the photodiode and amplification of the amplifier chain to convert from the measured microwave power at the VNA. To check this conversion, we measure the noise equivalent power of the photodiode in darkness, and compare to its specified value. 

To confirm that the modes result from the embedded magnetic material, we compare the optical measurement to a standard inductive ferromagnetic resonance (FMR). This is made via the reflected microwave power to the output port of the VNA, and is shown in Fig.\,\ref{BLS}(b). The change in microwave reflection coefficient $\Delta|S_{11}|$ with magnetic field is plotted over the same range as Fig.\,\ref{BLS}(a). We have confirmed that the resonances in Fig.\,\ref{BLS}(b) results from the YIG layer in FMR measurements over a wider magnetic field. The fact that the slope with magnetic field is the same in both measurements confirms that the optical signal results from Brillouin light scattering in the YIG. The band of resonances also has the same upper limit in both measurements.  The differences in the response -- in particular, that the microwave reflection spectra has more resonances than the optical BLS -- can be explained by the fact that, in the inductive measurement, the entire strip-line is probed, whereas the optical measurement is only sensitive to the region of the YIG film below the lens. The large number of resonances in the inductive measurement is due to strain inhomogeneity across the film from the film transfer process. 

The magnon modes observed in the optical measurement depend on overlap with both the microwave and optical fields \cite{haigh_selection_2018}. The Kittel mode has the correct symmetry to fulfill these requirements, and we tentatively assign the strongest scattering to this uniform mode. There are two other modes at higher frequency visible in Fig.\,\ref{BLS}(a). The mode spacing of these is too large to be due to perpendicular standing spin waves, given the thickness of the YIG film \cite{klingler_measurements_2014}. A possible candidate for these modes would be magneto-static surface spin waves \cite{damon_magnetostatic_1961} with wavevector set by the width of the microstrip antenna \cite{andrich_long-range_2017}. However, the robust identification of these modes requires further measurement and will be the subject of future work.

A fit to the Kittel mode in the BLS measurement gives a linewidth of $\Gamma\approx20$\,MHz, a value larger than is typical for high quality YIG thin films \cite{dubs_sub-micrometer_2017}. This is expected, because the current device has imperfections in the YIG layer due to the ion-implantation process, and strain disorder from the bonding process, such as the cracks visible in Fig.\,\ref{design}(c). These imperfections can be improved by further fabrication processes. Firstly, the damage from ion-implantation can be alleviated via annealing \cite{rachford_magnetization_1999}. Secondly, the strain disorder can be reduced by polishing the GGG from the back of the YIG/GGG bi-layer. It has also been possible to transfer a YIG layer crack-free.

The peak measured optical power of the BLS signal is $\approx$1.2\,nW. Given the local oscillator power $P_\textsc{lo}=65$\,$\mu$W and input microwave power 1\,mW, the total conversion efficiency is calculated to be $8\times10^{-16}$. This low value is to be expected, since the microwave coupling and magnon mode overlap in this devices have not been engineered. Therefore, to show the value of design we separate the coupling rate Eq.\,\ref{couplingrate} into an optical part $G_\text{opt}$ and the magnetic fill-factor $\eta_\text{mag}$, $G = G_\text{opt}\sqrt{\eta_\text{mag}}$, and estimate the obtained rate for the fabricated cavity.

The optical mode volume for a Gaussian beam can be estimated as \cite{dolan_femtoliter_2010}
\begin{equation}
V_\text{opt} = \pi w_1^2L/4,
\label{vol_est}
\end{equation}
where $L$ is the cavity optical path length and $w_1$ is the beam waist on the flat mirror surface. We estimate $w_1^2=(\lambda/\pi)\sqrt{\beta L}(1-L/\beta)$ in the parallax approximation, where $\beta$ is the radius of curvature of the lens. With the parameters of the measured device, $L\approx 12$\,$\mu$m and $\beta=70$\,$\mu$m, this yields $V_\text{opt} \approx 100$\,$\mu$m$^3$. This corresponding to $G_\text{opt}\approx50$\,kHz.

The magnon mode overlap in the device measured is poor. Taking the whole area of the cracked film part, we estimate $\eta\sim10^{-3}$, reducing the coupling rate to $G\approx50$\,Hz.

Compared to the whispering gallery mode $V_\text{opt} \approx 5000$\,$\mu$m$^3$, and the waveguide device of Ref.\,\onlinecite{zhu_waveguide_2020} $V_\text{opt} \approx 10^5$\,$\mu$m$^3$, the optical mode volume achieved here is a significant improvement. However, the microwave coupling and magnon confinement are lacking, severely limiting the conversion efficiency. The waveguide device \cite{zhu_waveguide_2020} has optimized magnon modes $\eta_\text{mag}\approx1$ and optimized microwave coupling through an microwave resonator, and even WGM mode devices ($\eta_\text{mag}\approx10^{-5}$) benefit from impedance matched microwave coupling to the Kittel mode \cite{osada_cavity_2016}.

\section{Conclusions}

We propose and demonstrate an open magneto-optical cavity device with optical mode volume limited by the thickness of an embedded magnetic layer. This design is tunable, has the correct polarized modes, and a significantly reduced optical mode volume compared to previous devices \cite{osada_cavity_2016,zhang_optomagnonic_2016,haigh_triple-resonant_2016,zhu_waveguide_2020}. 

We envisage that simple improvements in the demonstrated device design should enable the strong-coupling regime to be reached. By removing the GGG from the device via polishing, the cavity length can be reduced to ~3\,um, and using the lowest radius of curvature lens $\beta = 22$\,$\mu$m, the resulting mode volume would be $V_\text{opt}\approx7$\,$\mu$m$^3$ (from Eq.\,\ref{vol_est}). Combining this with lateral patterning of the YIG layer to confine a magnon mode to a disk with diameter 5\,$\mu$m, it should be possible to achieve $G=200$\,kHz using the open microcavity design. If we combine this with the discussed improvements in the magnon and optical linewidth to $\Gamma=1$\,MHz and $\kappa_{}=1$\,GHz, respectively, this would lead to a single photon cooperativity of $C = 4 G^2/\Gamma \kappa = 10^{-4}$. We would then require an optical pump power of $\approx	5$\,mW to achieve the strong coupling regime $\sqrt{n}G>\kappa_{},\Gamma$. In order to achieve cooling of magnetic mode via optical damping, $\Gamma_\text{opt} = 4nG^2/\kappa_{}$ \cite{aspelmeyer_cavity_2014,bittencourt_magnon_2019} comparable to the magnetic damping would require only $\approx 1$\,$\mu$W input power \cite{sharma_optical_2018}.

Finally, it will be necessary to couple microwaves efficiently into the resulting small volume of magnetic material. Elsewhere, we have demonstrated that this is possible using low impedance microwave resonators \cite{mckenzie-sell_low-impedance_2019}. With careful microwave circuit optimization it is possible to achieve coupling to femtolitre magnetic volumes \cite{ranjan_electron_2020}, to match that possible with open optical microcavities \cite{dolan_femtoliter_2010}.

As well as demonstrating progress towards microwave-optical conversion \cite{hisatomi_bidirectional_2016}, it is expected that the enhancement of the magnon-photon interaction demonstrated could have significant impact in magnonics \cite{kruglyak_magnonics_2010}, through the increased measurement sensitivity and in optical modification of the magnon dynamics. The versatility of the fabrication method means that antiferromagnetic materials could also be embedded in the microcavity in order to explore the interaction of optical photons with THz magnon modes \cite{walowski_perspective_2016,schlauderer_temporal_2019}.

\section*{acknowledgements}

We are grateful to Aurilien Trichet and Jason Smith (Oxford HiQ) for advice on open microcavities, Roger Webb (Surrey ion beam centre) for assistance with ion implantation, and Miguel Levy, Dries Van Thourhout, Koji Usami, Andreas Nunnenkamp and Paul Walker for useful discussions. This work was supported by the European Union's Horizon 2020 research and innovation programme under grant agreement No 732894 (FET Proactive HOT). The data plotted in the figures can be accessed at the Zenodo repository \footnote{\url{https://doi.org/10.5281/zenodo.4012308}}.

\bibliography{bibliography}

\end{document}